\documentclass[aps,11pt]{revtex4}
\usepackage{dcolumn}
\usepackage{graphicx}
\usepackage{amsmath}
\usepackage{amsfonts}
\usepackage{amssymb}
\usepackage{psfrag}
\usepackage{wrapfig}
\usepackage{subfigure}
\usepackage{makeidx}
\usepackage{bm}
\usepackage{epsf}
\usepackage{hyperref}
\usepackage{color}
\makeatletter

\newcommand{\Rmnum}[1]{\expandafter\@slowromancap\romannumeral #1@}

\makeatother

\begin{document}

\title{Phase transition and Quasinormal modes for Charged black holes in 4D Einstein-Gauss-Bonnet gravity}

\author{Ming Zhang$^{1}$\footnote{e-mail: mingzhang0807@126.com; zhangming@xaau.edu.cn}, Chao-Ming Zhang$^{2}$\footnote{e-mail: 843395448@qq.com}, De-Cheng Zou$^{2}$\footnote{e-mail: dczou@yzu.edu.cn}
and Rui-Hong Yue$^{2}$\footnote{e-mail: rhyue@yzu.edu.cn}}

\address{$^{1}$Faculty of Science, Xi'an Aeronautical University, Xi'an 710077 China\\
$^{2}$Center for Gravitation and Cosmology, College of Physical Science and Technology, Yangzhou University, Yangzhou 225009, China}

\date{\today}

\begin{abstract}
\indent
In the 4D EGB gravity, we consider the thermodynamic and phase transition of (charged) AdS black holes, for the negative GB coefficient $\alpha<0$ , the system allows two physical critical points corresponding to the reentrant phase transition when the charge $Q>2\sqrt{-\alpha}$. For arbitrary $\alpha>0$, the system always appears the Van der Waals phase transition. Then we study the quasinormal modes(QNMs) of massless scalar perturbations to probe the Van der Waals like small and large black holes(SBH/LBH) phase transition of (charged) AdS black holes. We find that the signature of this SBH/LBH phase transition in the isobaric process can be detected since the slopes of the QNMs frequencies change drastically different in small and large black holes near the critical point. The obtained results further support that the QNMs can be a dynamic probe of the thermodynamic properties in black holes.
\end{abstract}

\maketitle

\section{Introduction}
\label{intro}

During the past decades, higher order derivative curvature gravities, as the effective models of gravity in their low-energy limit string theories, have attracted considerable interest. Among these higher order derivative curvature gravities, the most vastly studied theory is the so-called Gauss-Bonnet gravity. In this theory, the Einstein's theory could be generalized in higher dimensions,
meanwhile excepting the linear dependence of the Riemann tensor all characteristics of usual general relativity can be kept \cite{Lanczos:1938sf,Lovelock:1971yv,Boulware:1985wk,Wheeler:1985qd}. Unfortunately, the Gauss-Bonnet(GB) term's variation is a total derivative
in 4 dimensional spacetime, which has no contribution to the gravitational dynamics. Therefore, one requires $D\geq 5$ for non-trivial gravitational dynamics. Recently, Glavan and Lin \cite{Glavan:2019inb} suggested a novel theory of gravity in 4-dimensional spacetime called ``4D Einstein Gauss-Bonnet gravity''(EGB). By rescaling the GB coupling constant $\alpha \to \alpha/(D-4)$ with $D$ the number of spacetime dimensions, and defining the 4-dimensional theory as the limit $D\to 4$, the GB term gives rise to non-trivial dynamics.
Furthermore, the spherically symmetric black hole solutions have been also constructed in this paper.
The generalization to other black holes has also appeared, for instance the charged AdS case\cite{Fernandes:2020rpa}, Lovelock\cite{Konoplya:2020qqh,Casalino:2020kbt}, rotating\cite{Kumar:2020owy,Ghosh:2020vpc}, Born-Infeld\cite{Yang:2020jno}, Bardeen\cite{Kumar:2020uyz}  Hayward\cite{Kumar:2020xvu}, etc. There also some important properties of the related 4D EGB black holes have been studied, such as the Spinning test particle in the black hole\cite{Zhang:2020qew}, the causality\cite{Ge:2020tid}, the stability and shadows\cite{Guo:2020zmf,Wei:2020ght}.

In the black hole physics, the thermodynamical phase transition of black hole is always a hot topic. Due to the AdS/CFT correspondence\cite{Maldacena:1997re,Gubser:1998bc,Witten:1998qj},
lots of attentions have been attracted to study the phase transition of black holes in anti-de Sitter(AdS) space mainly. Recently thermodynamics of AdS black holes has been studied in the extended phase space where the cosmological constant is treated as the pressure of the system \cite{Kastor:2009wy,Kubiznak:2012wp},
where it was found that a first order small and large black holes phase transition is allowed and the $P-V$ isotherms are analogous to the Van der Waals fluid.
More discussions including reentrant phase transitions and more general Van der Waals behavior in this direction can be found as well \cite{Gunasekaran:2012dq,Hendi:2012um,Zhao:2013oza,Zou:2013owa,
Dehghani:2014caa,Hennigar:2015esa,Zhang:2014jfa,Altamirano:2014tva,
Wei:2015iwa,Xu:2014kwa,Sadeghi:2016dvc,Hansen:2016ayo}.
On the other hand, the quasinormal modes (QNMs) of dynamical perturbations are considered as the characteristic sounds of black holes. The QNMs of the dynamical perturbations are expected to reflect the black hole phase transitions in their surrounding geometries through frequencies and damping times of the oscillations.
In fact, the thermodynamic phase transition of the AdS black hole in the dual field theory corresponds to the beginning of instability of a black hole. With lots of researches on this issue, more and more evidences of the connections between the QNMs of black holes and the thermodynamic phase transitions were found\cite{Gubser:2000mm}-\cite{Chabab:2016cem}.

Until now, this mentioned $P-V$ phase transition for various black holes in the 4D GB gravity were studied in \cite{HosseiniMansoori:2020yfj,Hegde:2020yrd,
Wang:2020pmb,Singh:2020mty,Li:2020vpo,Hegde:2020xlv,Wei:2020poh,
Singh:2020xju}. Among these research papers, the Van der Waals-like (SBH/LBH) phase transition for the black hole was found in 4D neutral AdS EGB black hole\cite{Li:2020vpo,Hegde:2020xlv}, the charged AdS\cite{Wei:2020poh} and the Bardeen AdS case\cite{Singh:2020xju}. Meanwhile, the QNMs of dynamical perturbation for the black hole was also reported in 4D neutral EGB black hole\cite{Konoplya:2020bxa,Churilova:2020aca}, the neutral dS\cite{Churilova:2020mif} and the regularized case\cite{Cuyubamba:2020moe}. Motivated by these results and the extensive importance of AdS/CFT correspondence, the aim of this paper is to study whether signature of Van der Waals like SBH/LBH phase transition of charged AdS black holes in 4D EGB gravity can be reflected by the dynamical QNMs behavior with the massless scalar perturbation.

The paper is organized as follows: in Sect.\ref{pv}, we firstly discuss the phase transition of (charged) AdS black holes in 4-dimensional EGB gravity in the extended phase space.
Then, we give discussions for QNM frequencies under test scalar field perturbations in Sect.\ref{2s}, and disclose the phase transition can be reflected by the QNM frequencies of dynamical perturbations. We end the paper with closing remakes in the last section.

\section{Thermodynamics and phase transition of Charged AdS black holes}
\label{pv}

As mentioned above, by rescaling the GB coupling parameter $\alpha \to \alpha/(D-4)$ and then taking the limit $D\to 4$, Glavan and Lin\cite{Glavan:2019inb} obtained a non-trivial 4D black hole solution.
Recently, the static and spherically symmetric charged AdS black hole solution in 4D Einstein-Gauss-Bonnet (EGB)
gravity is given by \cite{Fernandes:2020rpa}
\begin{eqnarray}
&&ds^2=-f(r)dt^2+\frac{1}{f(r)}dr^2+r^2d\Omega_2,\label{metric}\\
&&f(r)=1+\frac{r^2}{2\alpha}\left(1-\sqrt{1+4\alpha\left(\frac{2M}{r^3}
-\frac{Q^2}{r^4}-\frac{1}{l^2}\right)} \right),\label{solution}
\end{eqnarray}
where $M$ and $Q$ are the mass and charge of black hole, $\alpha$ is the Gauss-Bonnet coefficient with dimension $(length)^2$. When $\alpha\to 0$, $f(r)$ reduces to solution of RN AdS black hole in the general relativity. Moreover, Fernandes further investigate the related thermodynamic properties of this charged AdS black hole in this paper \cite{Fernandes:2020rpa}.

Now we reconsider the thermodynamics of this 4D charged AdS black hole in the so-called extended phase space.
In fact, in the extended phase space, the cosmological constant $\Lambda$
is usually regarded as the thermodynamic pressure $P=-\frac{\Lambda}{8\pi}$ in the geometric units $G_N=\hbar=c=k=1$.
Then, the mass $M$, Hawking temperature $T$ and entropy $S$ of 4D charged EGB-AdS black holes in the extended phase
space can be written as
\begin{eqnarray}
&&M=\frac{Q^2}{2r_+}+\frac{\alpha}{2r_+}+\frac{r_+}{2}+\frac{4\pi P}{3}r_+^3,\label{M}\\
&&T=\frac{2r_+^3}{r_+^2+\alpha}P-\frac{Q^2-r_+^2+\alpha}{4\pi r_+(r_+^2+2\alpha)},\\
&&S=\pi r_+^2+4\pi\alpha\ln{r_+}.\label{TS}
\end{eqnarray}

From the Hawking temperature (\ref{TS}), the equation of state can be obtained as
\begin{eqnarray}
P=(\frac{\alpha}{r_+^3}+\frac{1}{2r_+})T+\frac{Q^2+\alpha-r_+^2}{8\pi r_+^4} \label{eos}
\end{eqnarray}
As usual, a critical point is determined as the inflection point of $P$,
\begin{eqnarray}
\frac{\partial P}{\partial r_+}\Big|_{T=T_c, r_+=r_c}
=\frac{\partial^2 P}{\partial r_+^2}\Big|_{T=T_c, r_+=r_c}=0.\label{inflection}
\end{eqnarray}
Then we can obtain corresponding critical temperature and critical pressure
\begin{eqnarray}
&&T_c=\frac{r_c^2-2Q^2-2\alpha}{2\pi r_c(r_c^2+6\alpha)}, \label{critT} \\
&&P_c=-\frac{Q^2(3r_c^2+2\alpha)}{8\pi r_c^4(r_c^2+6\alpha)}+\frac{r_c^4-5\alpha r_c^2-2\alpha^2}{8\pi r_c^4(r_c^2+6\alpha)},\label{critvalue}
\end{eqnarray}
here the subscript ``c'' represents the critical values of the physical quantities. The critical horizon radius have two values
\begin{eqnarray}
&&r_{c1}=\sqrt{3(Q^2+2\alpha)+\sqrt{3(3Q^2+4\alpha)(Q^2+4\alpha)}},\label{critrc} \\
&&r_{c2}=\sqrt{3(Q^2+2\alpha)-\sqrt{3(3Q^2+4\alpha)(Q^2+4\alpha)}}.
\end{eqnarray}

When the GB coefficient $\alpha$ is positive, the critical radius $r_{c2}$ is always  imaginary and the critical radius $r_{c1}$ is always positive. From Eq.\ref{critT} to Eq.\ref{critrc}, we can verify that the critical temperature and pressure stay positive value ($T_{c1}>0$ and $P_{c1}>0$) for arbitrary parameters $\alpha>0$. The system admit only one physical critical point ($r_{c1}$, $T_{c1}$ and $P_{c1}$), therefore we can tell the system always allow a Van der Waals phase transition when $\alpha$ takes the positive values. Now we consider the negative GB coefficient ($\alpha<0$). We find that the system allows two positive critical point ($r_{c1}, T_{c1}, P_{c1}$ and $r_{c2}, T_{c2}, P_{c2}$) when the charge $Q$ and GB coefficient $\alpha$ satisfy certain constraint $Q>2\sqrt{-\alpha}$. For example, when $\alpha=-0.01, Q=0.21$, we can obtain $r_{c1}=0.3256, T_{c1}=0.4017, P_{c1}=0.2459$ and $r_{c2}=0.1965, T_{c2}=1.1205, P_{c2}=1.2539$. The so-called reentrant phase transition (RPT) will appear under this constraint.
If we take the constraint $0<Q\leq 2\sqrt{-\alpha}$, the system has not any physical critical point. In this paper, our main aim is to study whether signature of Van der Waals like SBH/LBH phase transition of charged AdS black holes in 4D EGB gravity can be reflected by the dynamical QNMs behavior with the massless scalar perturbation. As mentioned above, when $\alpha$ takes the negative value, the system could only allow the reentrant phase transition, which is hard to be reflected by the dynamical QNMs behavior. Therefore we only consider the positive GB coefficient ($\alpha>0$) in the subsequent paragraph.

\begin{figure}[htb]
\centering
\subfigure[$Q=0.1$ and $T_c=0.219$]{\label{fig:subfig:11} 
\includegraphics[width=2.2in]{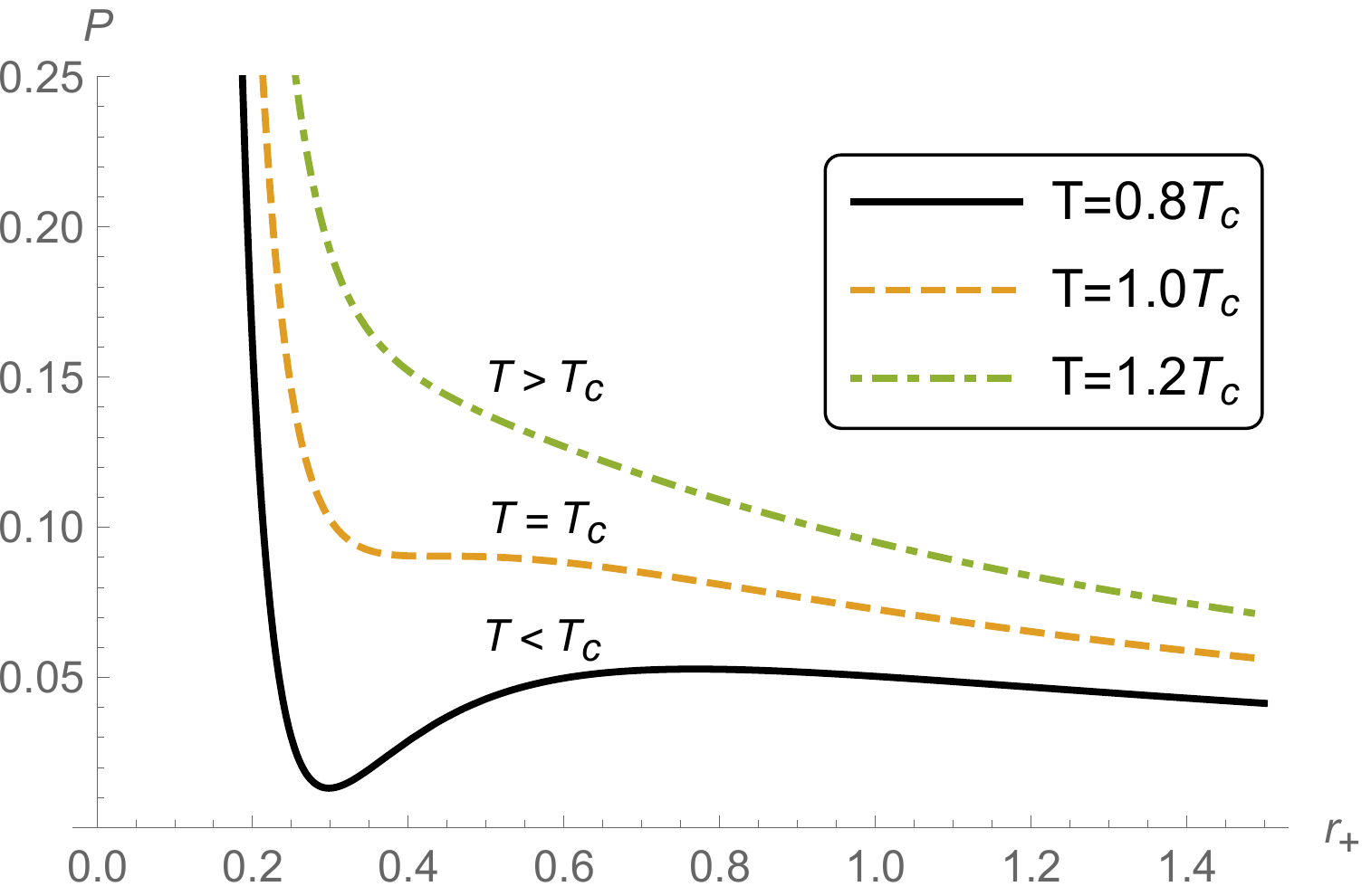}}
\hfill
\subfigure[$Q=0$ and $T_c=0.2556$]{\label{fig:subfig:12} 
\includegraphics[width=2.2in]{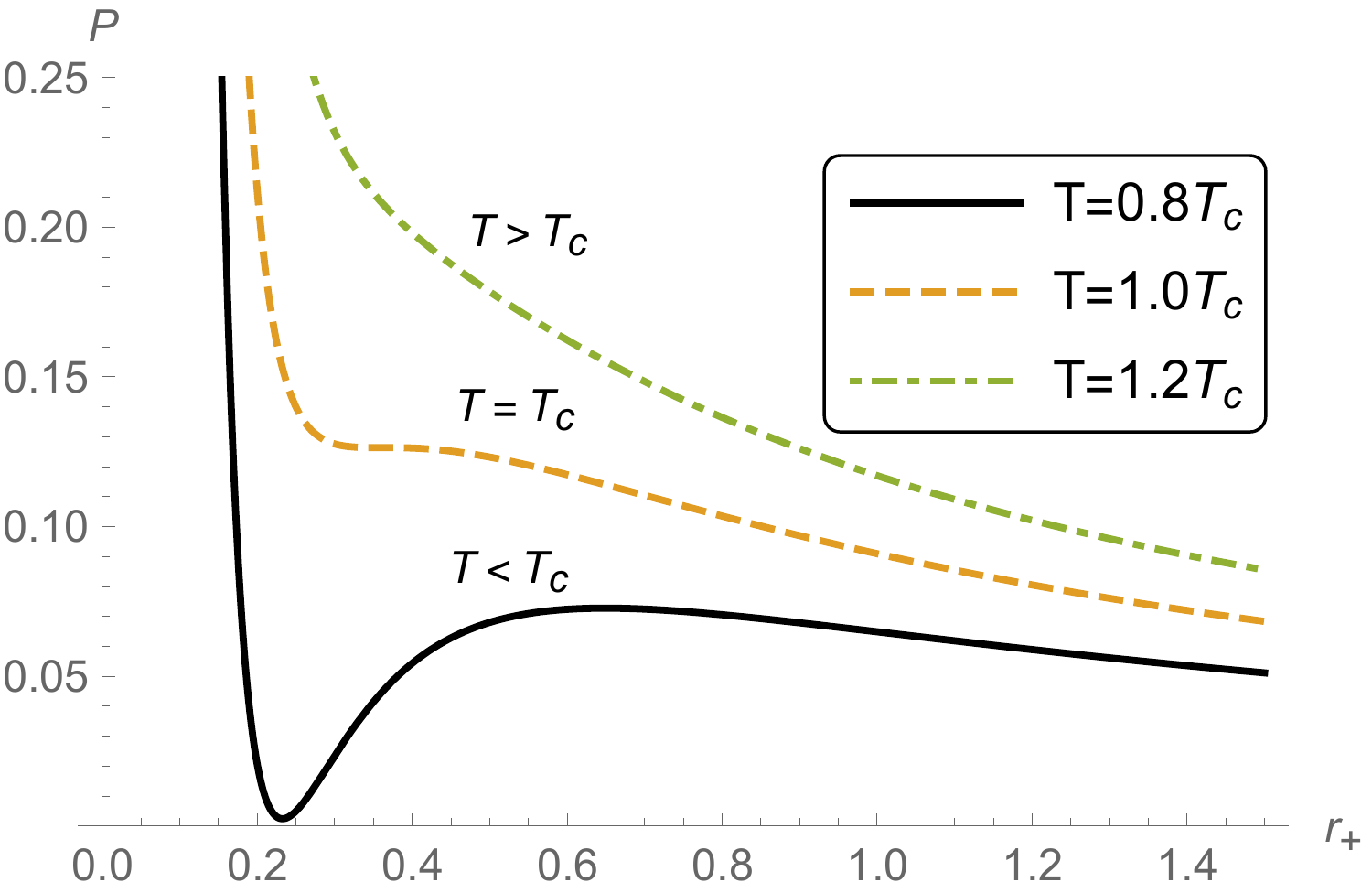}}
\hfill
\caption{The $P-r_+$ diagram of charged and uncharged AdS black holes with $\alpha=0.01$.}\label{fig11}
\end{figure}

For instance, we can obtain a critical point with $r_c=0.4387$, $T_c=0.219$ and $P_c=0.0904$ by fixed $\alpha=0.01$ and $Q=0.1$.
Moreover, in the uncharged case, Eq.(\ref{critT})-Eq.(\ref{critrc}) can be written as
\begin{eqnarray}
  &&T_c=\frac{1+\sqrt{3}}{2(3+\sqrt{3})\sqrt{6+4\sqrt{3}}\pi\sqrt{\alpha}}, \\
  &&P_c=\frac{13+7\sqrt{3}}{3168\pi\alpha+1824\sqrt{3}\pi\alpha}, \\
  &&r_c=\sqrt{2}\sqrt{3\alpha+2\sqrt{3}\alpha}.
\end{eqnarray}
For example, we can get a critical point with $r_c=0.3596$, $T_c=0.2556$ and $P_c=0.1264$ by fixed $\alpha=0.01$.
We plot the $P-r_+$ isotherm diagram around the critical temperature $T_c$ for this charged and uncharged AdS black hole, see Fig.\ref{fig11}. The dotted line with $T>T_c$ corresponds to the ``idea gas'' phase behavior, and when $T<T_c$ the Van der Waals like small/large black hole phase transition will appear.

The thermodynamic phase transition is importantly determined by the behavior of Gibbs free energy $G$ which obeys the following thermodynamic relation $G=M-TS$ with
\begin{eqnarray}
&&G=2\pi r_+^3 P\left(\frac{2}{3}-\frac{r_+^2+4\alpha\ln{r_+}}{r_+^2+2\alpha}\right)+\frac{Q^2+r_+^2+\alpha}{2r_+} \nonumber\\
&&+\frac{(Q^2-r_+^2+\alpha)(r_+^2+4\alpha\ln{r_+})}{4r_+(r_+^2+2\alpha)}.\label{free}
\end{eqnarray}
Here $r_+=r_+(P,T)$ is understood as a function of the pressure and temperature, via the equation of state (\ref{eos}).

\begin{figure}[htb]
\centering
\label{fig:subfig:12} 
\includegraphics[width=2.3in]{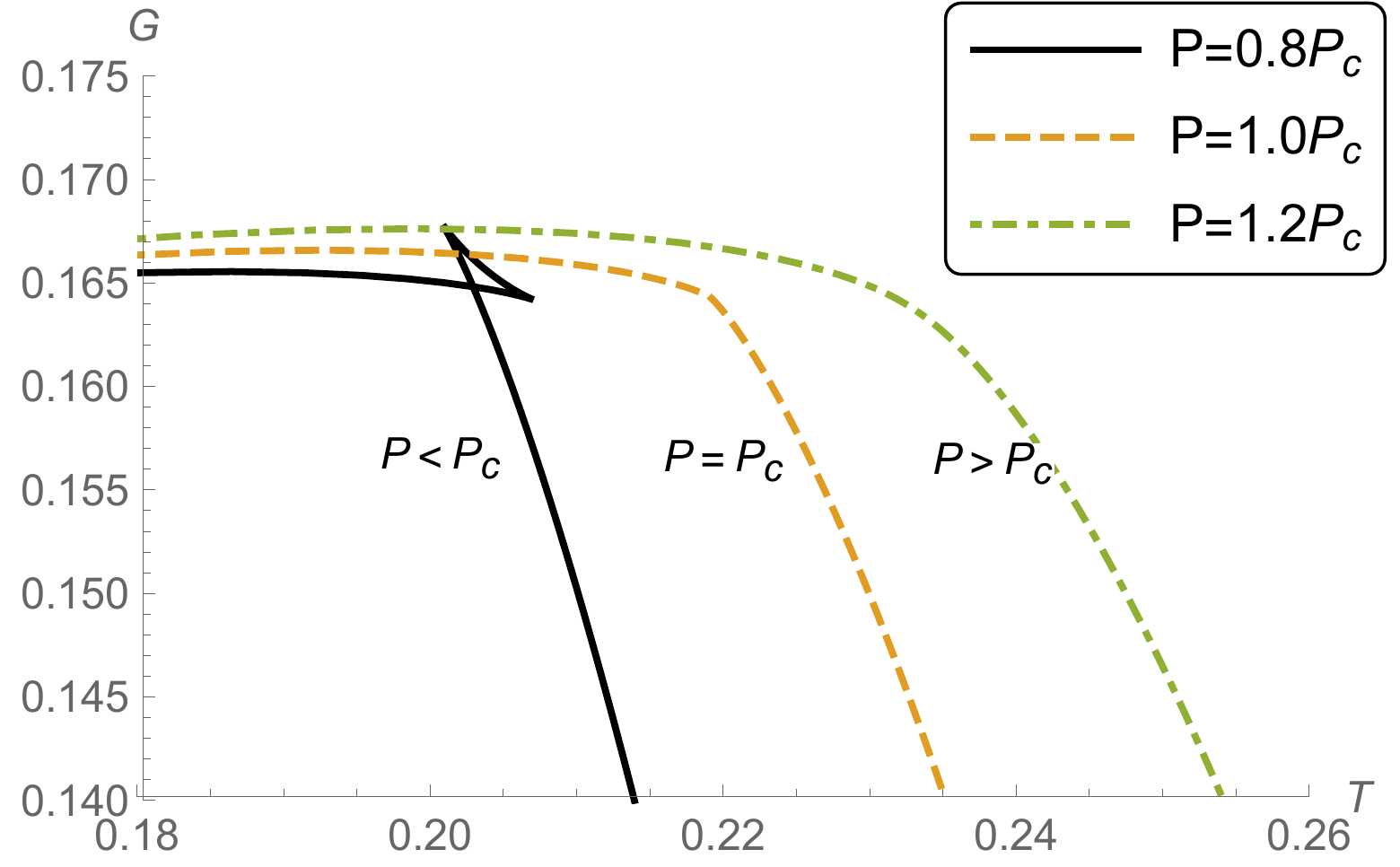}
\hfill%
\label{fig:subfig:13} 
\includegraphics[width=2.3in]{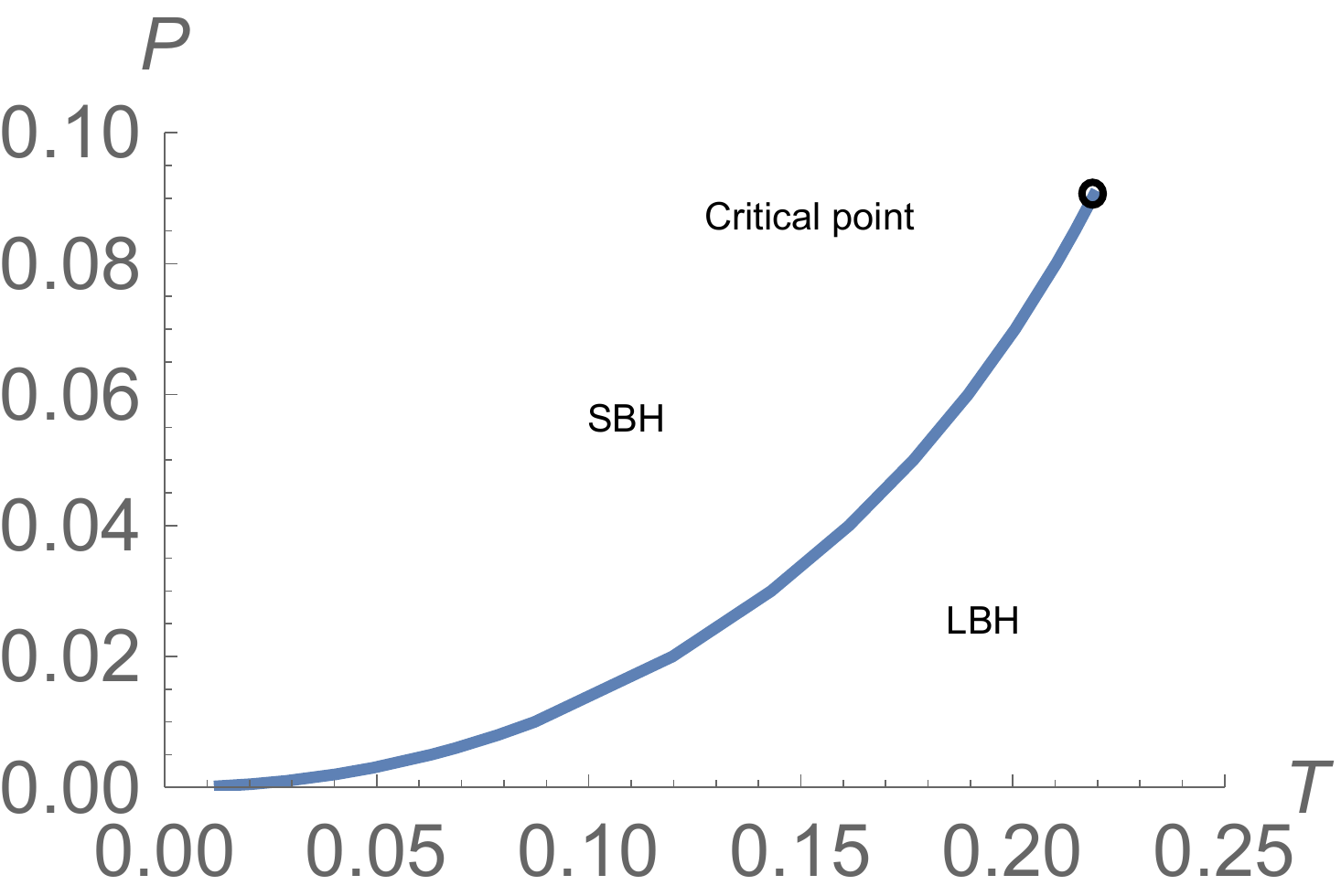}
\caption{The Gibbs free energy $G$ (left panel) and coexistence line of small/large black holes phase transition (right panel) with $\alpha=0.01$ and $Q=0.1$.}\label{fig12}
\end{figure}

\begin{figure}[htb]
\centering
\label{fig:subfig:121} 
\includegraphics[width=2.5in]{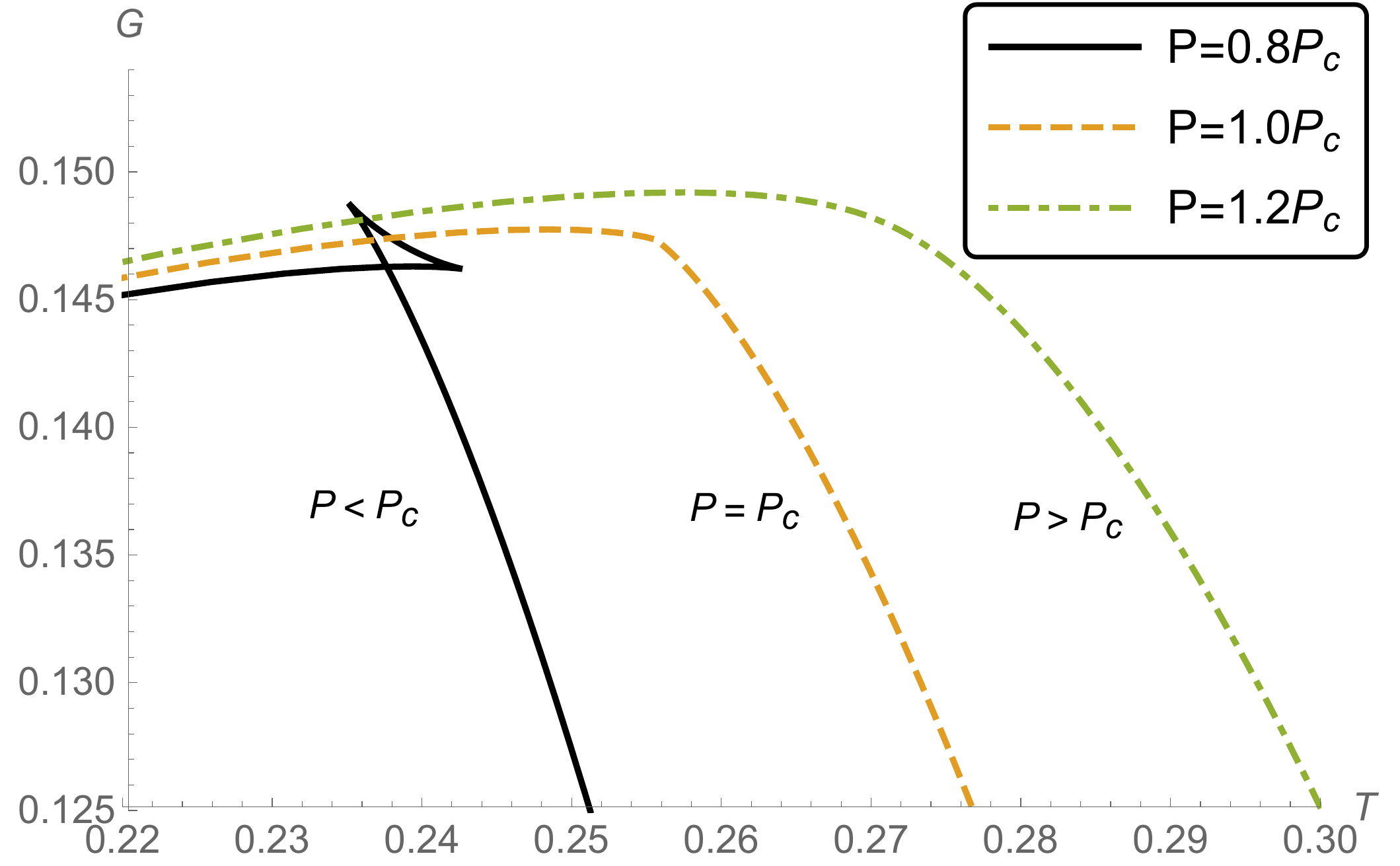}
\hfill%
\label{fig:subfig:131} 
\includegraphics[width=2.4in]{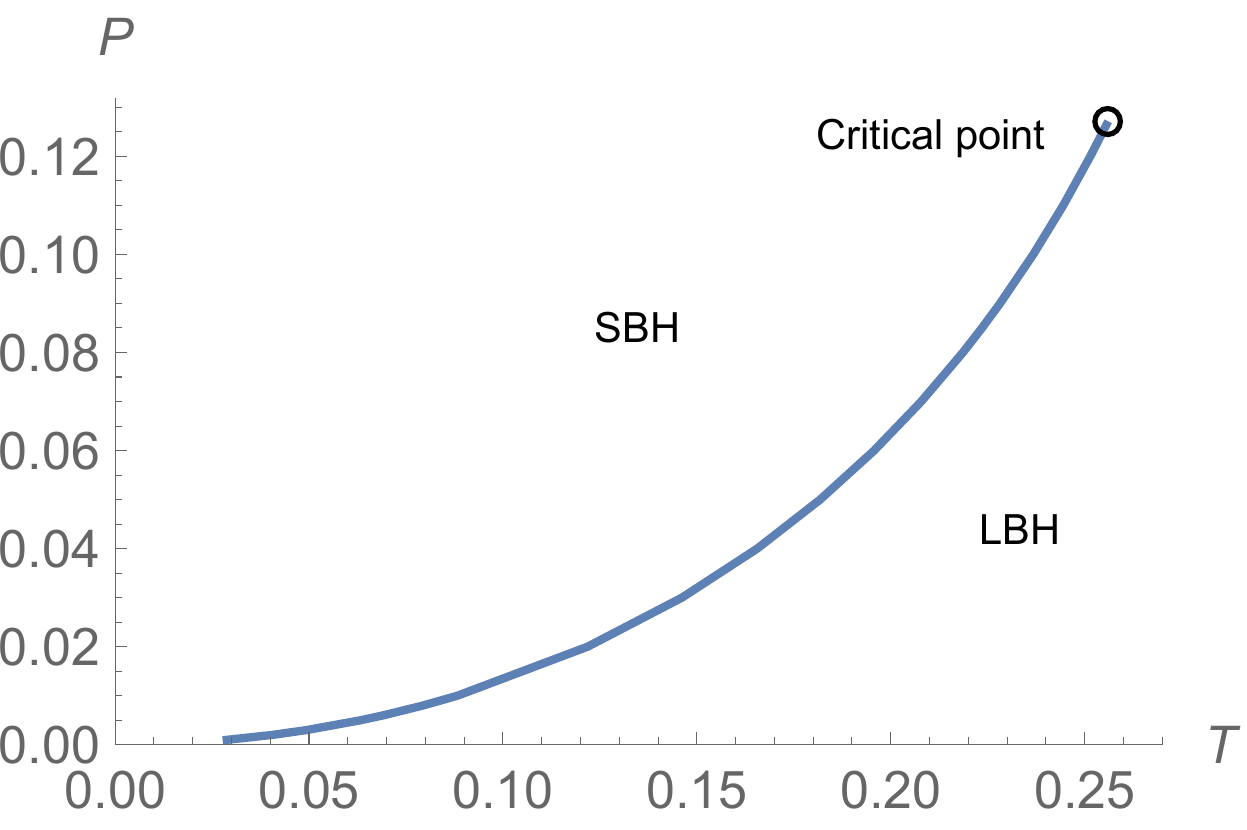}
\caption{The Gibbs free energy $G$ (left panel) and coexistence line of small/large uncharged black holes phase transition (right panel) with $\alpha=0.01$ and $Q=0$.}\label{fig121}
\end{figure}

In the left panel of Fig.\ref{fig12}, the swallow tail behavior of the Gibbs free energy $G$ shows that the system contains a Van der Waals like first order phase transition. The coexistence line in the ($P,T$) plane is plotted in the right panel of Fig.\ref{fig12}, the point in this curve represents the small and large black hole possess the same Gibbs free energy and temperature. We can find that the coexistence line is very similar to that of the Van der Waals fluid. At the end of the coexistence line the small circle means the critical point. When $T<T_c$, the small-large black hole phase transition occurs.

For the Van der Waals liquid-gas system, the liquid-gas structure will undergo a second order phase transition, which does not suddenly change at the critical point $(V=V_c, T=T_c, P=P_c)$. This phenomenon can be described by the Ehrenfest's description \cite{Linder,Stanley}, which contains the first and second Ehrenfest's equations \cite{sNieuwenhuizen,Zemansky}
\begin{eqnarray}
&&\frac{\partial P}{\partial T}\Big|_S=\frac{C_{P2}-C_{P1}}{TV(\zeta_2-\zeta_1)}
=\frac{\Delta C_P}{TV\Delta\zeta},\label{eq:38a}\\
&&\frac{\partial P}{\partial T}\Big|_V=\frac{\zeta_2-\zeta_1}{\kappa_{T2}-\kappa_{T1}}
=\frac{\Delta\zeta}{\Delta\kappa_{T}}.\label{eq:39a}
\end{eqnarray}
In a genuine second order phase transition, the two equations have to be satisfied simultaneously.
Here $\kappa_T$ and $\zeta$ represent the isothermal compressibility coefficients
and the volume expansion of the system respectively
\begin{eqnarray}
\zeta=\frac{1}{V}\frac{\partial V}{\partial T}\Big|_P,\quad
\kappa_T=-\frac{1}{V}\frac{\partial V}{\partial P}\Big|_T. \label{eq:40a}
\end{eqnarray}
Following the method in Ref.\cite{Zou:2014mha}, we find that the Ehrenfest’s equations is satisfied, which means in the 4-dimensional charged/neutral EGB-AdS black hole, this phase transition at the critical point is of the second order.
It has the same nature with the liquid-gas phase transition at the critical point.

\section{Perturbation of AdS black hole in 4D Einstein-Gauss-Bonnet gravity}
\label{2s}

In order to reflect the thermodynamical stabilities in dynamical perturbations, we can study the
evolution of a massless scalar field perturbation around this 4-dimensional EGB-AdS black hole.

A massless scalar field $\Phi(r,t,\Omega)=\phi(r)e^{-i\omega t}Y_{lm}(\Omega)$, obeys Klein-Gordon equation
\begin{eqnarray}
\nabla^2_{\mu}\Phi(t,r,\Omega)=\frac{1}{\sqrt{-g}}\partial_\mu\left(\sqrt{-g}g^{\mu\nu}\partial_\nu\Phi(t,r,\Omega)\right)=0.\label{KG}
\end{eqnarray}
Then radial equation for the function $\phi(r)$ is obtained as
\begin{eqnarray}
\phi''(r)+\frac{f'(r)}{f(r)}\phi'(r)+\left(\frac{\omega^2}{f(r)^2}-\frac{l(l+1)}{r^2 f(r)}
-\frac{f'(r)}{rf(r)} \right)\phi(r)=0,\label{KG1}
\end{eqnarray}
where $\omega$ are complex numbers $\omega=\omega_r + i\omega_{im}$, corresponding to the QNM frequencies of the oscillations describing the perturbation.

Near the horizon $r_+$, we can impose the boundary condition of the scalar field, $\phi(r)\to (r-r_+)^\frac{i\omega}{4\pi T}$.Then we define $\phi(r)$ as $\varphi(r)exp[-i\int\frac{\omega}{f(r)}dr]$,
where the $exp[-i\int\frac{\omega}{f(r)}dr]$ asymptotically approaches to ingoing wave near horizon, we can rewrite Eq.~(\ref{KG1}) into
\begin{eqnarray}
\varphi''(r)+\left(\frac{f'(r)}{f(r)}-\frac{2i\omega}{f(r)} \right)\varphi'(r)-\left(\frac{f'(r)}{r f(r)}+\frac{l(l+1)}{r^2 f(r)} \right)\varphi(r)=0.\label{omegaKG1}
\end{eqnarray}
For Eq.~(\ref{omegaKG1}), we have $\varphi(r)=1$ in the limit of $r\rightarrow r_+$. At the AdS boundary $(r\rightarrow\infty)$, we need $\varphi(r)=0$. Under these boundary conditions, we will numerically solve Eq.~(\ref{omegaKG1}) to find QNM frequencies by adopting the shooting method.

\begin{figure}[htb]
\centering
\label{fig:subfig:a} 
\includegraphics[width=2.3in]{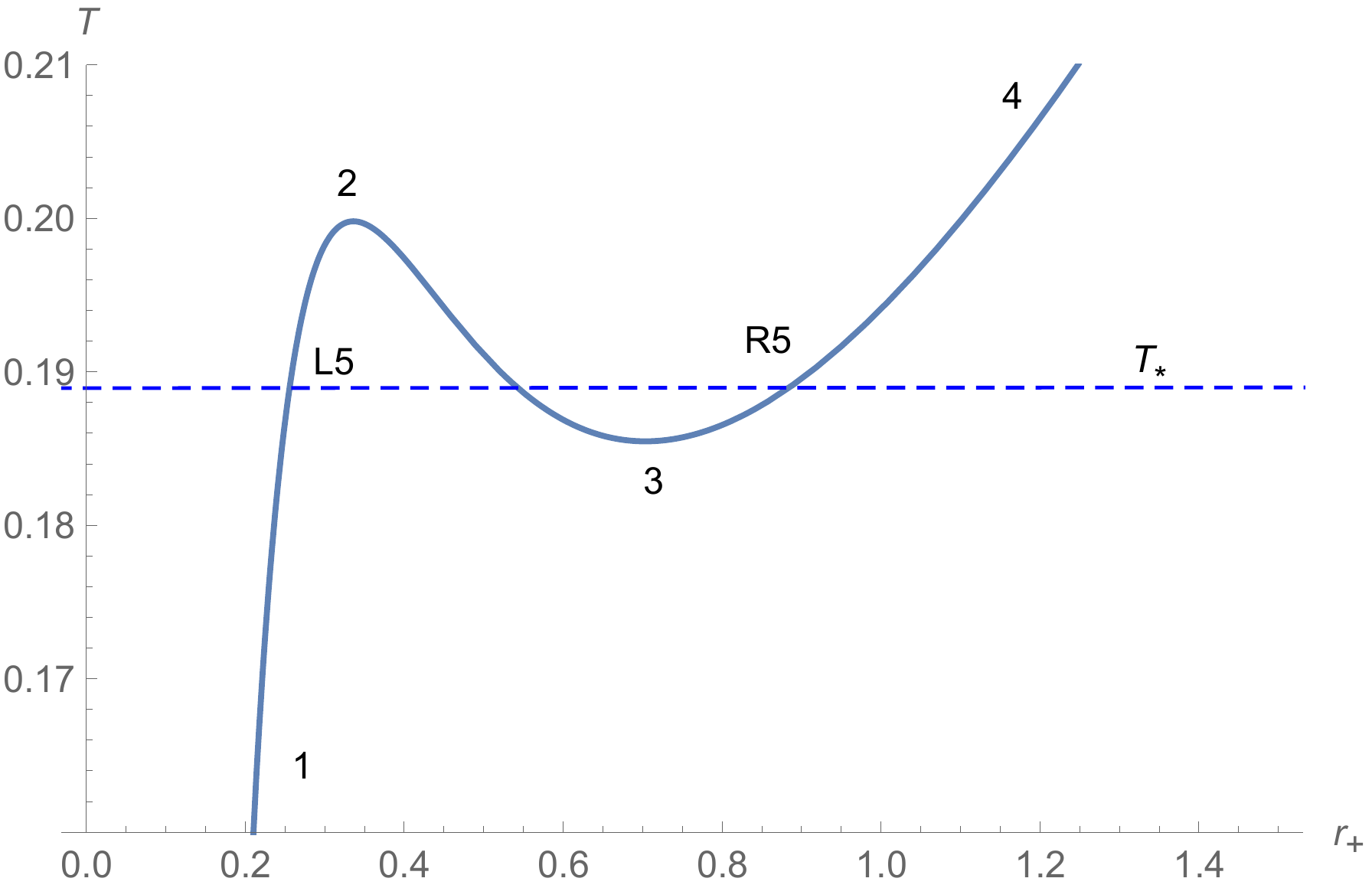}
\hfill%
\label{fig:subfig:b} 
\includegraphics[width=2.3in]{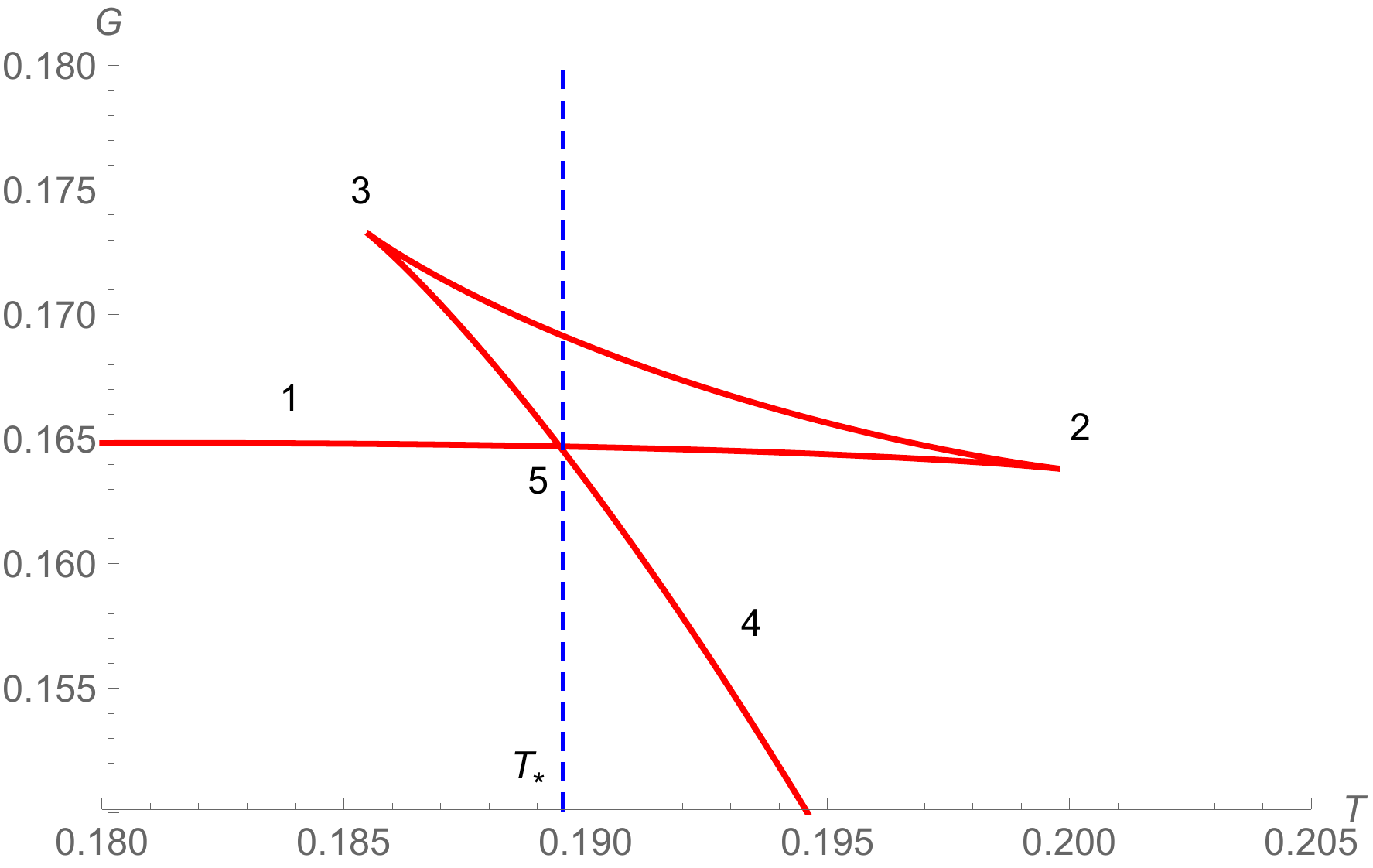}
\caption{$T-r_+$ (upper panel) and $G-T$(left panel) diagrams of 4D EGB AdS black holes with $\alpha=0.01, Q=0.1$ and $P\simeq0.06$.}\label{fig1}
\end{figure}

In the upper panel of Fig.~\ref{fig1}, we plot $T-r_+$ diagram of charged AdS black holes with fixed pressure $P=0.06<P_c=0.219$ in four dimensional EGB gravity. When the pressure $P<P_c$, there is an inflection point and the behavior of the system is similar with the Van der Waals system. The critical point can be obtained from
\begin{eqnarray}
\frac{\partial T}{\partial r_+}\Big|_{P=P_c, r_+=r_c}
=\frac{\partial^2 T}{\partial r_+^2}\Big|_{P=P_c, r_+=r_c}=0.
\end{eqnarray}
The right panel of Fig.~\ref{fig1} shows the behavior of the Gibbs free energy.
In this figure we mark the cross point ``5" between the solid line as ``1-5" and
and the solid line as ``4-5". The cross point means the Gibbs free energy $G$ and pressure $P$ coincide for small and large black holes. In the upper panel of Fig.~\ref{fig1}, we separated the point ``5" into ``L5" and ``R5" for the same Gibbs free energy and the chosen $T_*\approx0.18946$, which represented the small and large black hole can coexist. Moreover, between points ``1-5" or ``1-L5" the marked physical phase corresponds to the small black hole, while between points ``5-4" or ``R5-4" the indicated physical phase denotes the large black hole.

\begin{figure}[htb]
\centering
\label{fig:subfig:a} 
\includegraphics[width=2.3in]{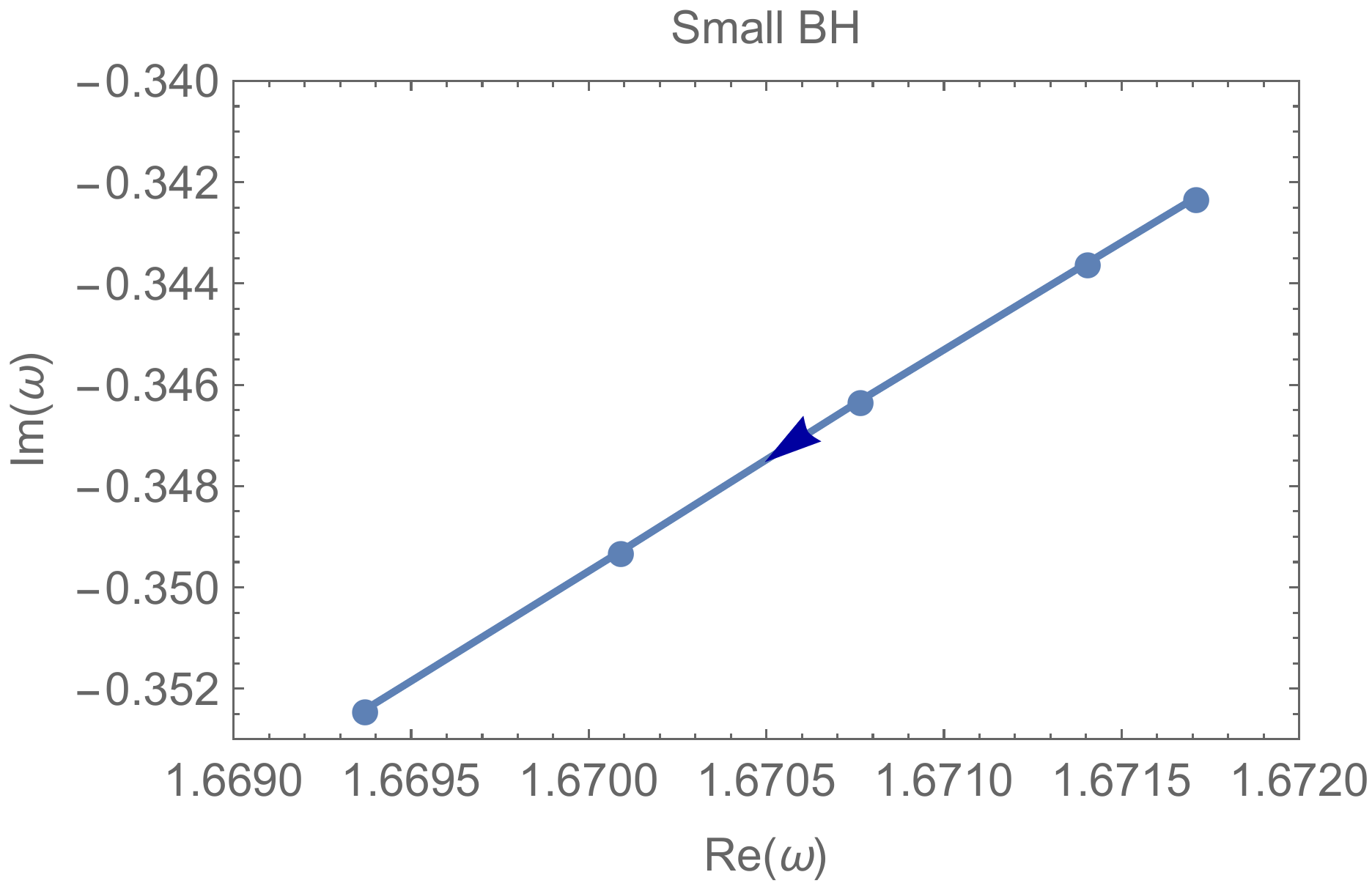}
\hfill%
\includegraphics[width=2.3in]{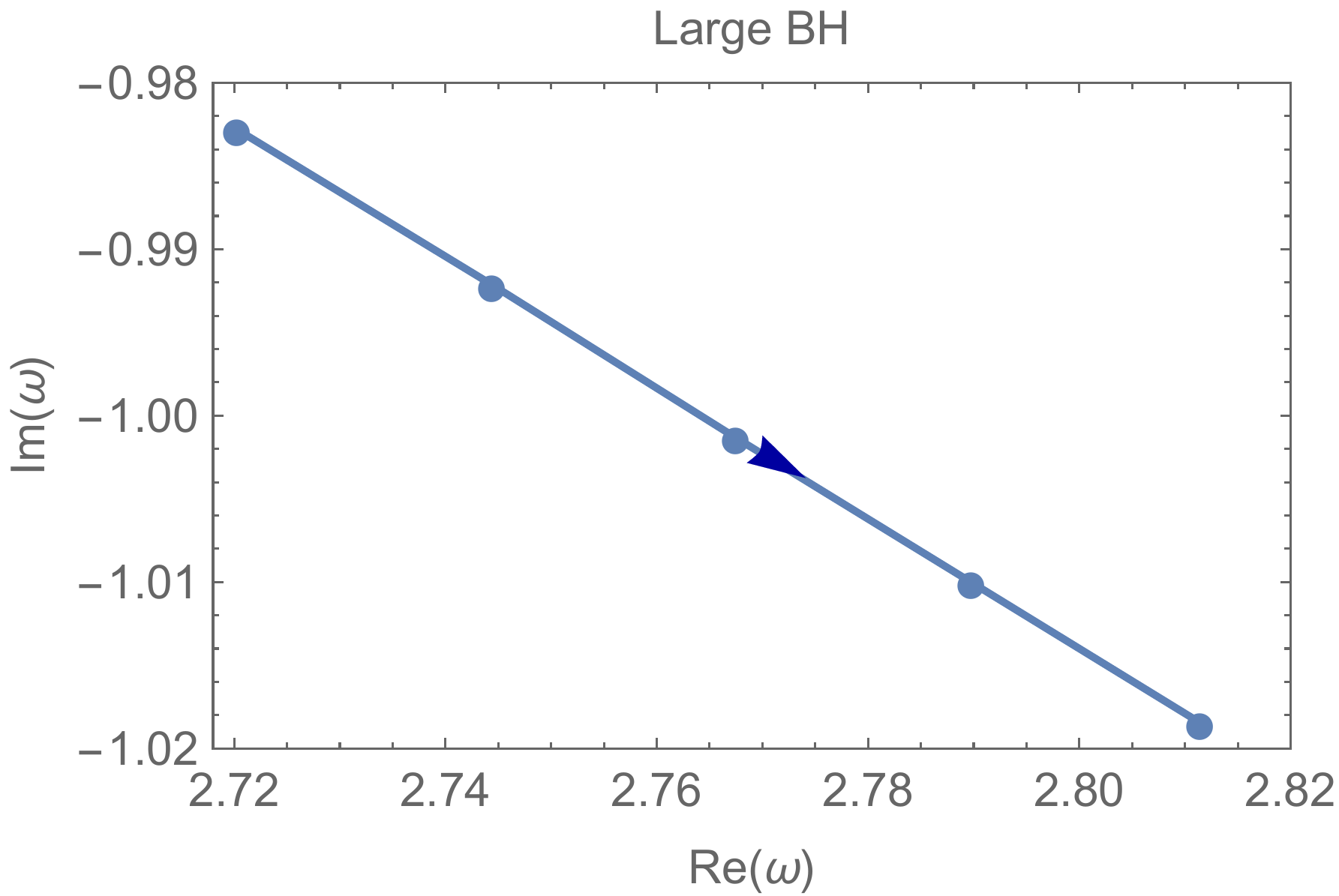}
\caption{The behavior of QNMs for large and small black holes in the
complex-$\omega$ with $Q=0.1$ and $l=0$. The arrow indicates the increase of black hole horizon.}\label{fig2}
\end{figure}

\begin{figure}[htb]
\centering
\label{fig:subfig:a1} 
\includegraphics[width=2.3in]{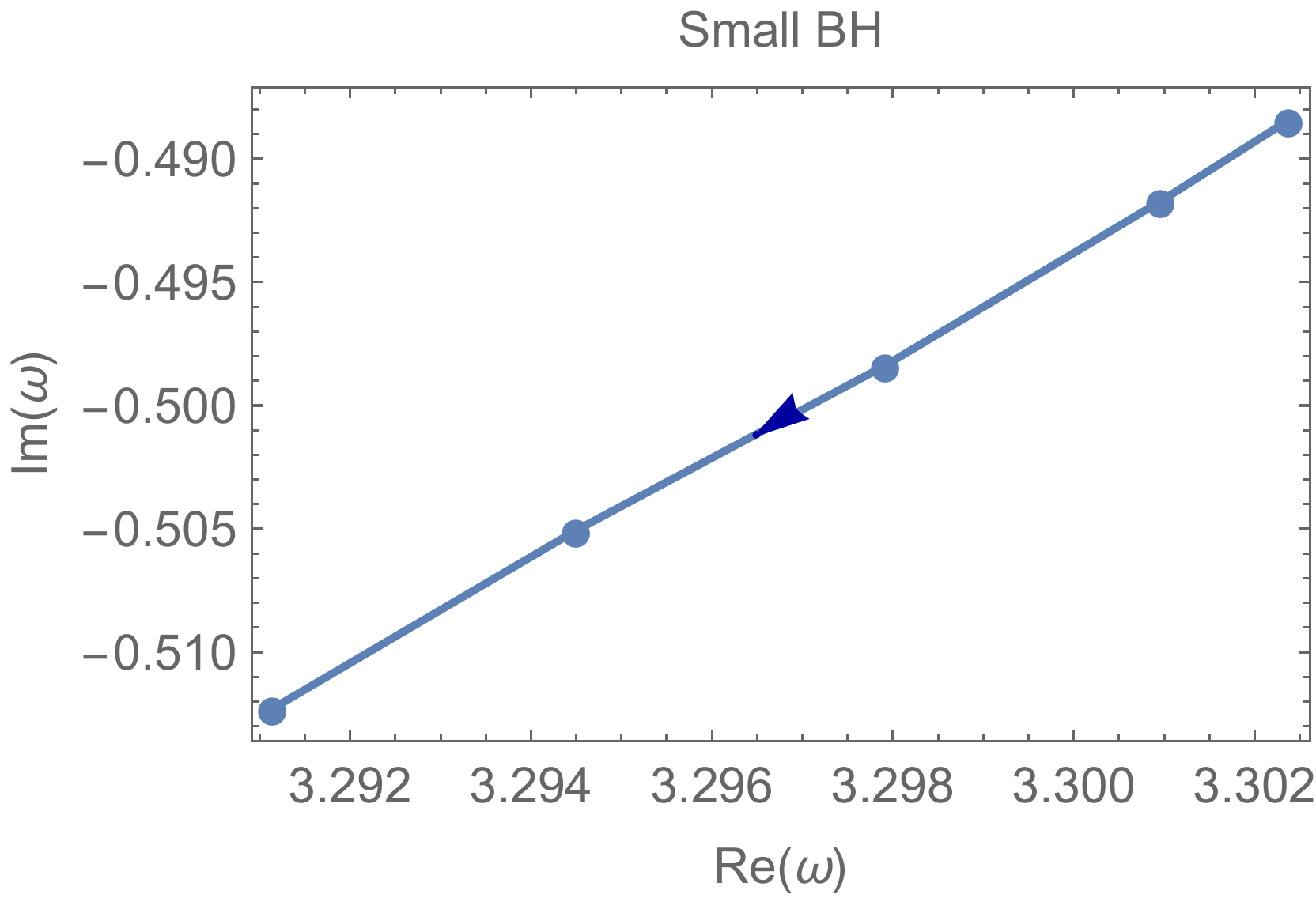}
\hfill%
\includegraphics[width=2.3in]{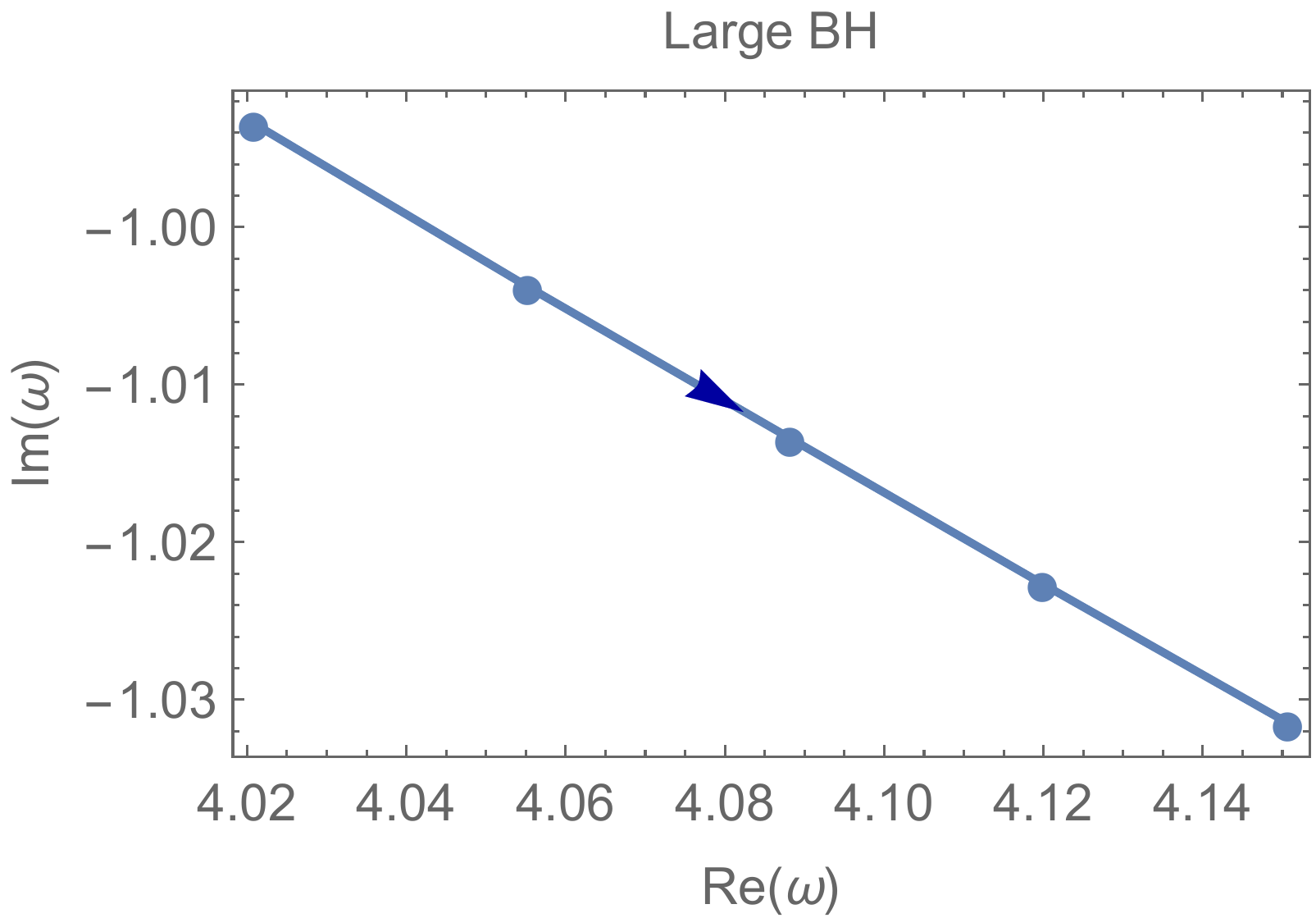}
\caption{The behavior of QNMs for large and small black holes in the
complex-$\omega$ with $Q=0.1$ and $l=1$. The arrow indicates the increase of black hole horizon.}\label{fig3}
\end{figure}

TABLE.\ref{table1} lists the QNM frequencies of massless scalar perturbation ($l=0$ and $1$) for small and large charged black holes near the SBH/LBH phase transition point. When the temperature decreases from phase transition temperature $T_*$, the radius of black hole becomes smaller and smaller, which corresponding to small black hole phase. We can find that the absolute values of imaginary part of QNM frequencies decrease in this process, meanwhile the real part of the frequencies change very little. On the other hand, the black hole will get larger, when temperature for the large black hole phase increases from the phase transition temperature $T_*$. We can also see that both the real part and the absolute value of imaginary part increase, which means that despite the massless scalar perturbation outside the black hole oscillates even more but it decays faster. These results are similar with the discussions reported in \cite{Liu:2014gvf,Chabab:2016cem}.
The Fig.~\ref{fig2} and Fig.~\ref{fig3} respectively
illustrates the QNM frequencies with $l=0$ and $l=1$ for small and large black hole phases. The arrows indicated the increasing direction of the black hole's size.

\begin{figure}[htb]
\subfigure[$l=0$]{\label{fig:subfig:b} 
\includegraphics[width=2.3in]{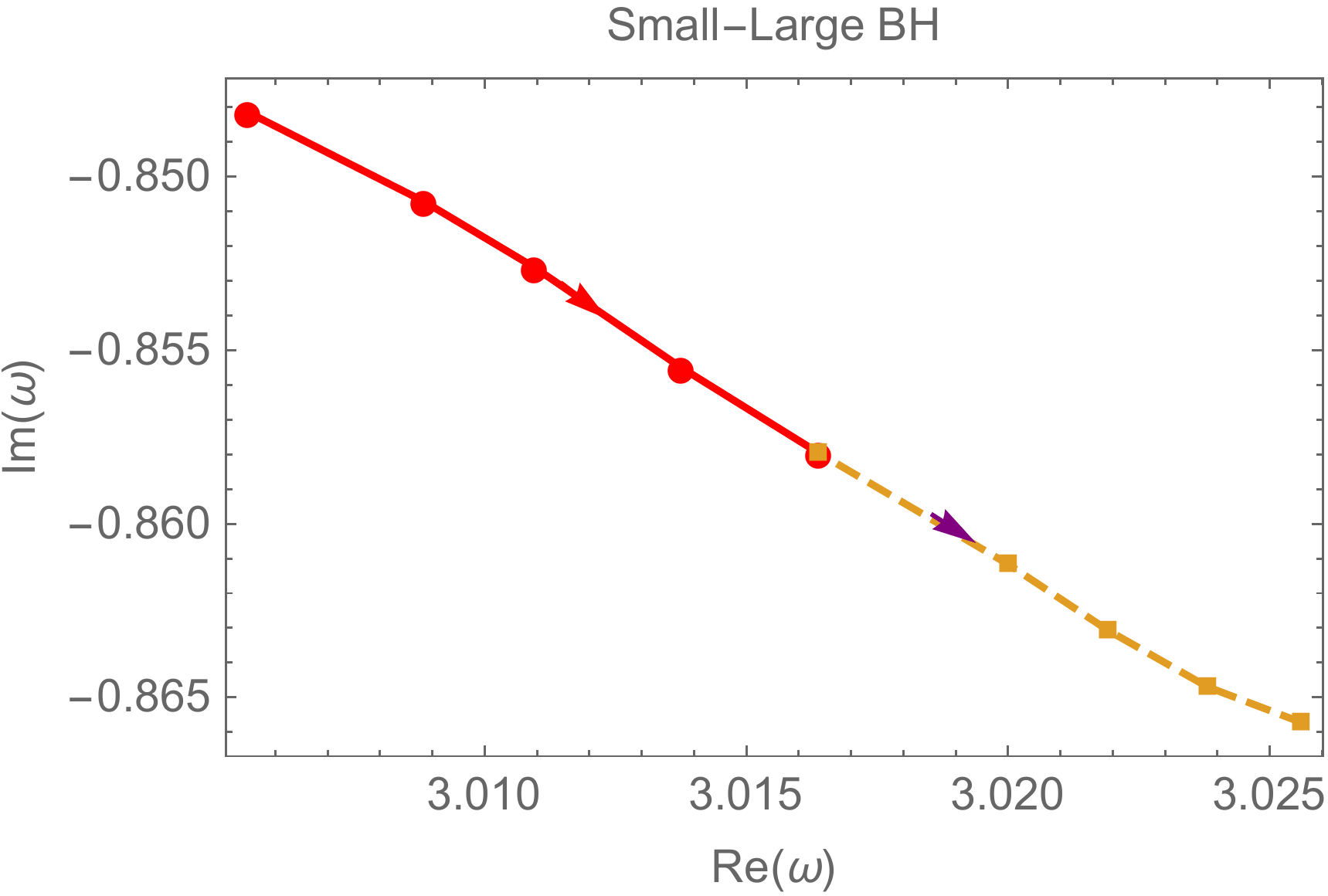}}
\hfill%
\subfigure[$l=1$]{
\includegraphics[width=2.3in]{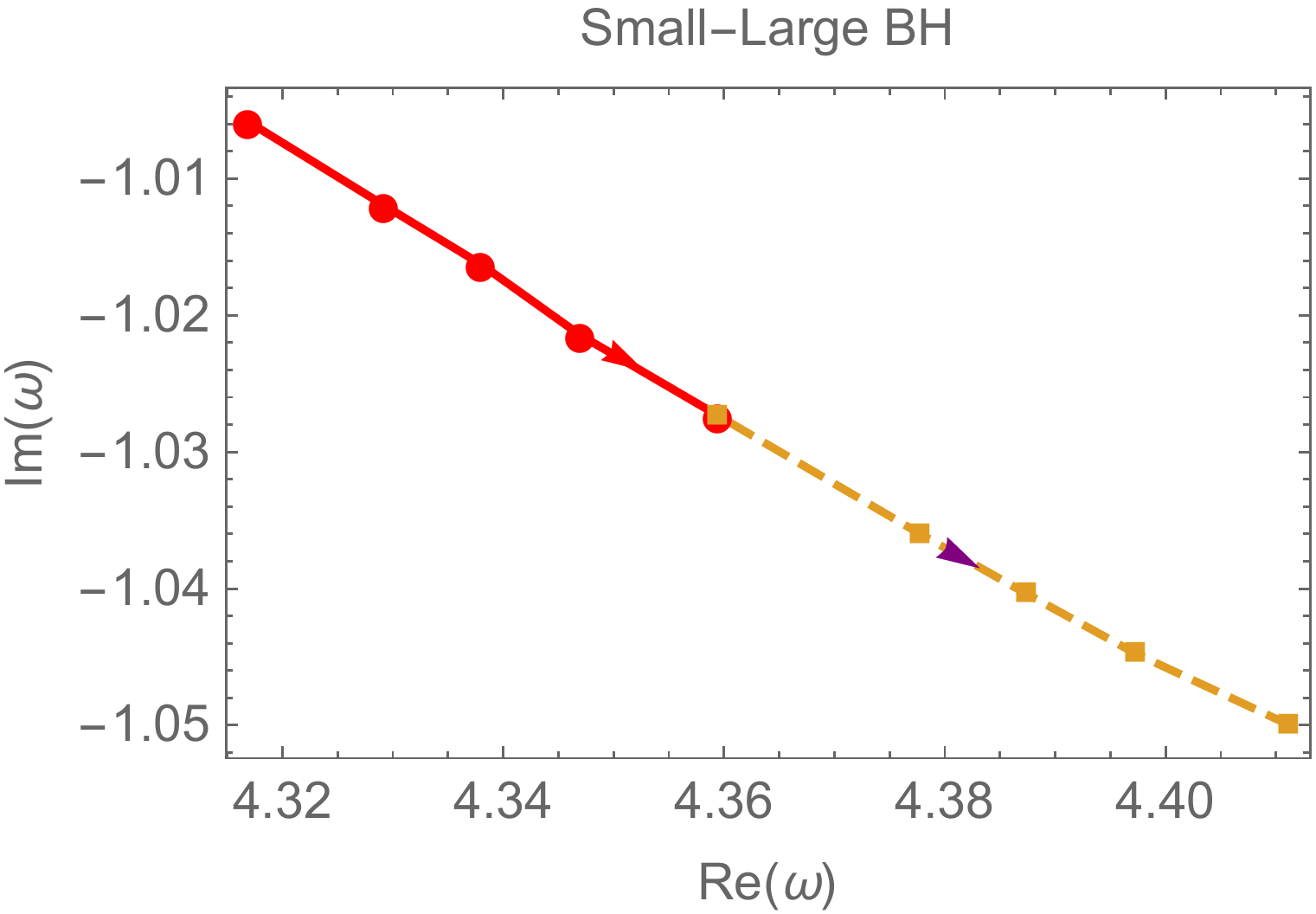}}
\caption{The behavior of QNM frequencies for large (dashed) and small (solid)
black holes in the complex-$\omega$ with $Q=0.1$.
The arrow indicates the increase of black hole horizon.}\label{fig4}
\end{figure}

Moreover, at the critical position $P=P_c$, with $P_c\simeq0.06$, a second-order phase transition occurs. The QNM frequencies of the small and large black hole phases (for $l=0$ and $l=1$) are exhibited in Fig.~\ref{fig4}.
We can see at the critical point the QNM frequencies of this two black hole
phases possess the same behavior when the black hole horizon increases.

\begin{table}
\caption{The QNM frequencies of massless scalar perturbation with the
change of black hole temperature with $\alpha=0.01$ and $Q=0.1$. The upper part, above the horizontal line, is for the small black hole phase, while the lower part is for the large black hole phase.}
\label{table1}
\begin{center}
\begin{tabular}{lrcc}
\hline
\hline
$T$    & $r_+$   & $\omega (l=0)$    & $\omega (l=1)$ \\ \hline
0.1855 & 0.24719 & 1.67171-0.342286I & 3.30238-0.488428I\\
0.186  & 0.24833 & 1.67141-0.343580I & 3.30097-0.491701I\\
0.187  & 0.25072 & 1.67077-0.346294I & 3.29793-0.498354I\\
0.188  & 0.25325 & 1.67009-0.349273I & 3.29450-0.505050I\\
0.189  & 0.25594 & 1.66937-0.352404I & 3.29115-0.512268I\\ \hline
0.190  & 0.91129 & 2.72028-0.982813I & 4.02092-0.993442I\\
0.191  & 0.93516 & 2.74446-0.992160I & 4.05526-1.003787I\\
0.192  & 0.95733 & 2.76751-1.001339I & 4.08825-1.013420I\\
0.193  & 0.97819 & 2.78981-1.010033I & 4.11994-1.022668I\\
0.194  & 0.99800 & 2.81150-1.018480I & 4.15075-1.031497I\\ \hline\hline
\end{tabular}
\end{center}
\end{table}

\begin{figure}[htb]
\centering
\label{fig:subfig:a2} 
\includegraphics[width=2.3in]{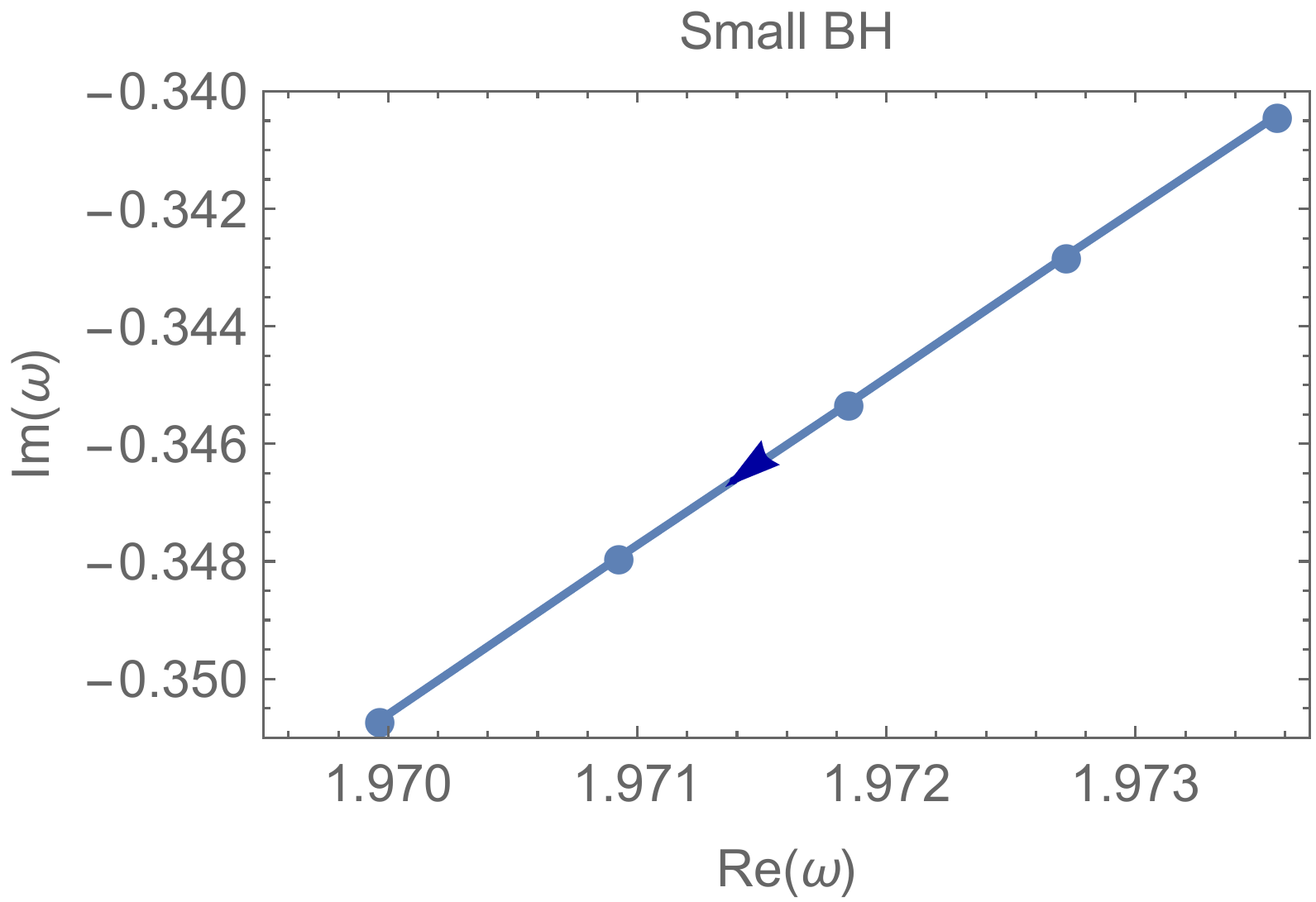}
\hfill%
\includegraphics[width=2.3in]{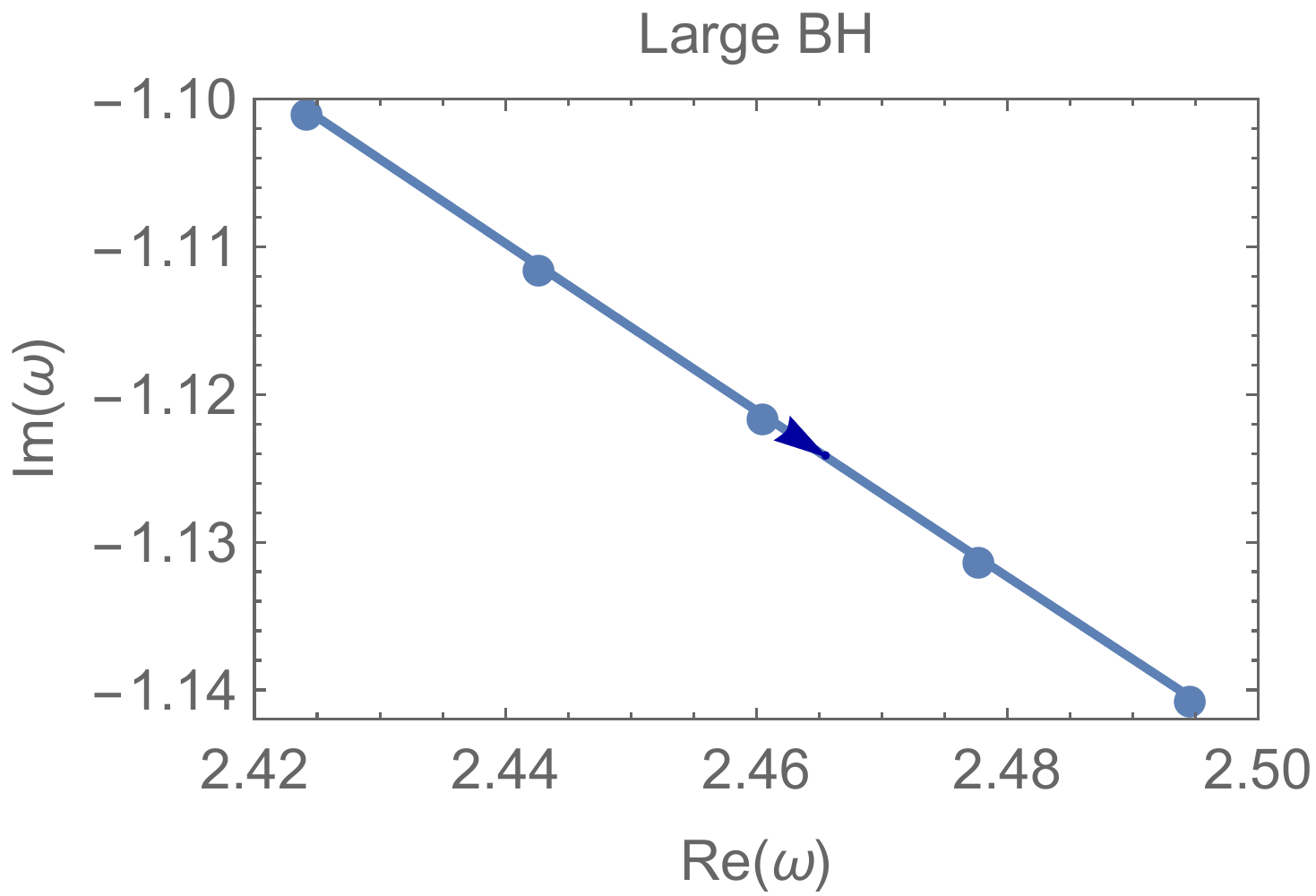}
\caption{The behavior of QNMs for large and small black holes in the
complex-$\omega$ with $Q=0$. The arrow indicates the increase of black hole horizon.$Q=0, l=0$}\label{fig5}
\end{figure}

\begin{figure}[htb]
\centering
\label{fig:subfig:a3} 
\includegraphics[width=2.3in]{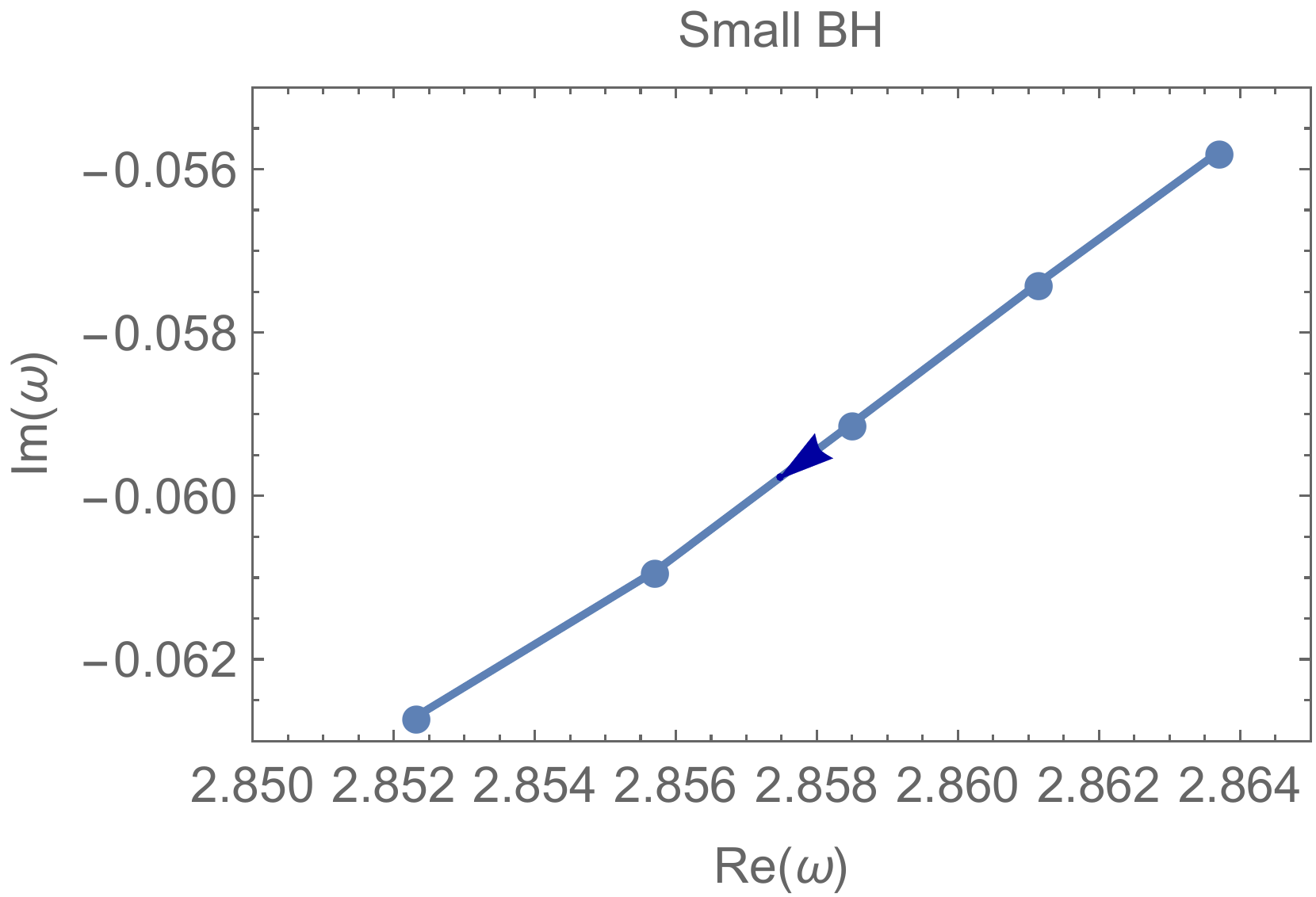}
\hfill%
\includegraphics[width=2.3in]{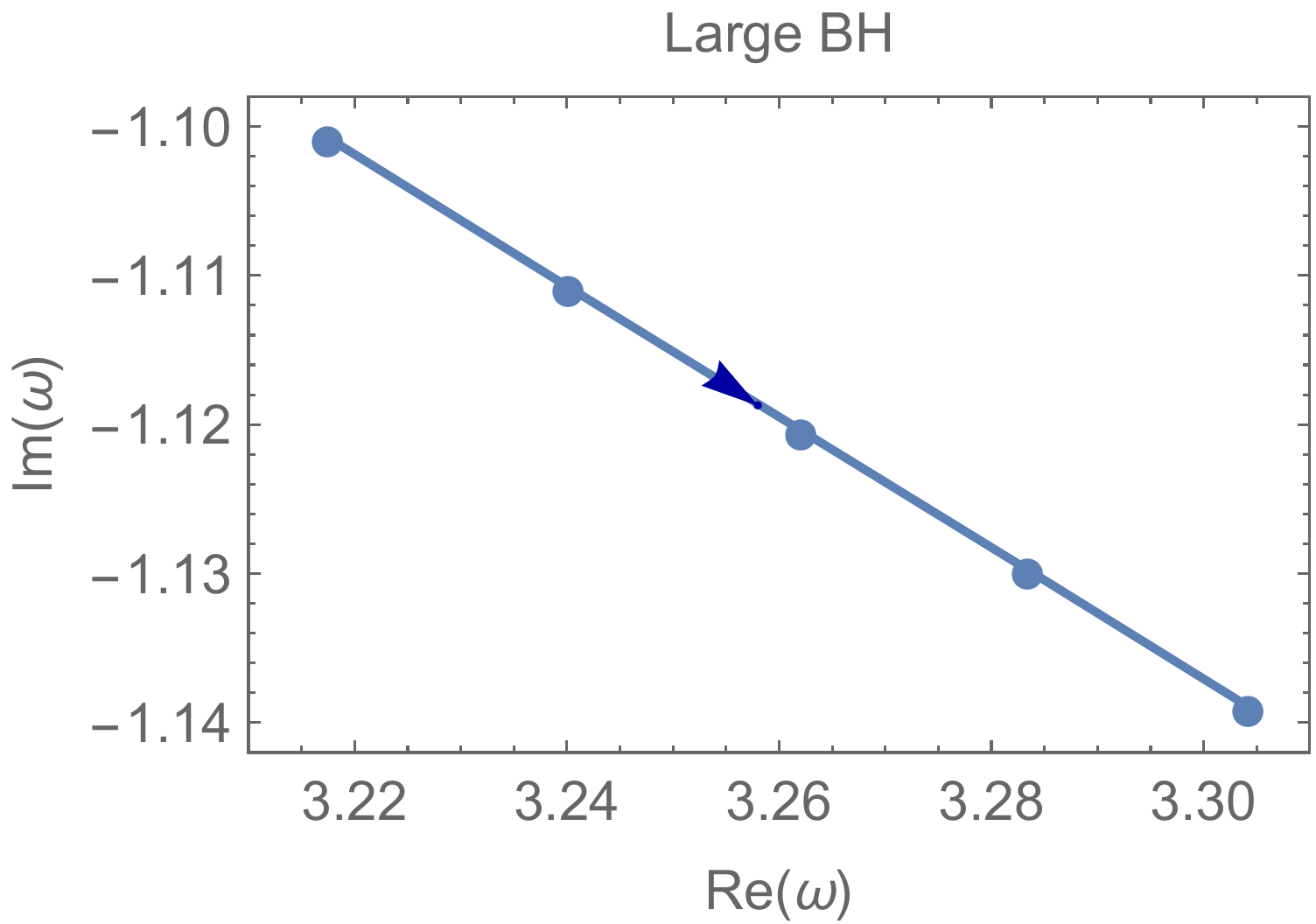}
\caption{The behavior of QNMs for large and small black holes in the
complex-$\omega$ with $Q=0$. The arrow indicates the increase of black hole horizon.$Q=0, l=1$}\label{fig6}
\end{figure}

In the neutral case, we can obtain similar $T-r_+$ and $G-T$ diagrams to the charged case. For instance, the coexistence temperature $T_*$ equals to $0.21860$ when taking pressure $P=0.08<P_c=0.1264$. The QNM frequencies of massless scalar perturbation (for $l=0$ and $1$) around small and large uncharged  black holes for first order SBH/LBH phase transition are listed in TABLE.\ref{table2}, which also exhibit a similar behavior to that of the charged case. The QNM frequencies with $l=0$ and $l=1$ for small and large uncharged black hole phases are shown in Fig.~\ref{fig5} and Fig.~\ref{fig6}. Moreover, at the critical position, the corresponding  QNM frequencies of the small and large uncharged black hole phases are also qualitatively similar to the charged case.

\begin{table}
\caption{The QNM frequencies of massless scalar perturbation with the
change of black hole temperature with $\alpha=0.01$ and $Q=0$. The upper part,
above the horizontal line, is for the small black hole phase,
while the lower part is for the large black hole phase.}
\label{table2}
\begin{center}
\begin{tabular}{lrcc}
\hline
\hline
$T$    & $r_+$   & $\omega (l=0)$   & $\omega (l=1)$ \\ \hline
0.214 & 0.18938 & 1.97357-0.340401I & 2.86372-0.0557742I\\
0.215 & 0.19088 & 1.97273-0.342787I & 2.86116-0.0573858I\\
0.216 & 0.19245 & 1.97185-0.345297I & 2.85852-0.0591034I\\
0.217 & 0.19408 & 1.97093-0.347917I & 2.85572-0.0609143I\\
0.218 & 0.19579 & 1.96997-0.350686I & 2.85233-0.0626936I\\ \hline
0.219 & 0.79874 & 2.42426-1.100808I & 3.21754-1.100784I\\
0.220 & 0.81586 & 2.44275-1.111345I & 3.24017-1.110794I\\
0.221 & 0.83201 & 2.46057-1.121425I & 3.26214-1.120424I\\
0.222 & 0.84739 & 2.47780-1.131115I & 3.28349-1.129773I\\
0.223 & 0.86211 & 2.49461-1.140502I & 3.30429-1.138959I\\ \hline\hline
\end{tabular}
\end{center}
\end{table}

\section{Closing remarks}
\label{3s}

In the 4-dimensional Einstein Gauss-Bonnet gravity, we have studied the $P-V$ criticality and phase transition of AdS black holes in the extended phase space. We demonstrate the system allows two physical critical points corresponding to the reentrant phase transition when the GB coefficient is negative and the charge satisfies the constraint $Q>2\sqrt{-\alpha}$. For arbitrary positive parameter $\alpha$ and $Q$, the VdW-like SBH/LBH phase transition always happens both in the charged and neutral cases. Then, we further calculated the QNMs of massless scalar perturbations under 4 situations (charged/uncharged and $l=0$ or $1$). These results reveal that the slopes of the QNM frequency change drastically different in the small and large black hole phases as increasing of the horizon radius $r_+$, when the Van der Waals analogy SBH/LBH phase transition happens in the extended space. This clearly demonstrates the signature of the phase transition between small and large black holes. In addition, at the critical isobaric phase transitions, the QNM frequencies for both small and large black holes share the same behavior, which showing that QNMs are not appropriate to probe the black hole phase transition in the second order.

This is one more example exhibits that the QNM can provide the dynamical physical phenomenon of the thermodynamic phase transition of black holes in 4D EGB gravity. Since the QNM is expected to be detected and has strong astrophysical interest. The ability of QNMs to reflect the thermodynamic phase transition is interesting, which is expected to disclose the observational signature of the thermodynamic phase transition.

This work is supported by the National Natural Science Foundation of China under Grant Nos.11605152, 11675139 and 51802247, and Outstanding youth teacher programme from Yangzhou University.

\end{document}